\documentstyle{mn}
\input{psfig}

\begin{document}


\def\ergsec{\hbox{erg s$^{-1}$}}
\def\ergcmsec{\hbox{erg cm$^{-2}$ s$^{-1}$}}
\def\countsec{\hbox{counts s$^{-1}$}}
\def\photcmsec{\hbox{photon cm${-2}$ s$^{-1}$}}
\def\degmark{^\circ}
\def \rsun {\ifmmode$R$_{\odot}\else R$_{\odot}$\fi}
\def \nh {N$_{\rm H}$}
\def\ax{$\alpha_{\rm x}$}
\def \hcm {\hbox {\ifmmode $ H atoms cm$^{-2}\else H atoms cm$^{-2}$\fi}}
\def\approxgt{\mathrel{\hbox{\rlap{\lower.55ex \hbox {$\sim$}}
        \kern-.3em \raise.4ex \hbox{$>$}}}}
\def\approxlt{\mathrel{\hbox{\rlap{\lower.55ex \hbox {$\sim$}}
        \kern-.3em \raise.4ex \hbox{$<$}}}}
\def\la{\mathrel{\hbox{\rlap{\hbox{\lower4pt\hbox{$\sim$}}}\hbox{$<$}}}}
\def\ga{\mathrel{\hbox{\rlap{\hbox{\lower4pt\hbox{$\sim$}}}\hbox{$>$}}}}
\def\kms{${\rm km~s^{-1}}$}
\def\PL{${\rm P_{1.4GHz}}$}      
\def\Lel{${\rm L_{e.l.}}$}
\def\uv{{\it uv\ }}
\def\FR{Fanaroff-Riley}
\newcommand {\rosat} {{ROSAT }}
\newcommand {\einstein} {{\it Einstein }}
\newcommand {\exosat} {{EXOSAT }}
\newcommand {\asca} {{\it ASCA }}
\newcommand {\sax} {{\it BeppoSAX }}
\newcommand {\ginga} {{\it GINGA }}
\newcommand {\ie} {{\it  i.e.}}
\newcommand {\cf} {{\it  cf }}
\newcommand {\eg} {{e.g.}}
\newcommand {\etal} {et~al. }
\newcommand {\Msun} {{ M$_{\odot}$}}
\newcommand {\degree} {$^{\circ}$}
\newcommand {\sqcm} {cm$^{2}$}
\newcommand {\cbcm} {cm$^{3}$}
\newcommand {\persqcm} {cm$^{-2}$}
\newcommand {\percbcm} {cm$^{-3}$}
\newcommand {\s} {s$^{-1}$}
\newcommand {\gps} {g s$^{-1}$}
\newcommand {\kmps} {km s$^{-1}$}
\newcommand {\lts} {{\it lt}-s}
\newcommand {\yr} {yr$^{-1}$}
\newcommand {\cps} {counts~s$^{-1}$}
\newcommand {\ergs} {erg~s$^{-1}$}
\newcommand {\ergcms} {erg cm$^{-2}$ s$^{-1}~$}
\newcommand {\chisq} {$\chi ^{2}$}
\newcommand {\rchisq} {$\chi_{\nu} ^{2}$}
\newcommand {\cpskeV} {counts s$ ^{-1}$ keV$ ^{-1}$ }
\newcommand{\DXDYCZ}[3]{\left( \frac{ \partial #1 }{ \partial #2 }
                        \right)_{#3}}
\title[\sax Observations of 2-Jy Lobe-dominated
Broad-Line Sources]
{\sax Observations of 2-Jy Lobe-dominated Broad-Line Sources.
I. The Discovery of a Hard X-ray Component}  

\author[P. Padovani et al.]{Paolo Padovani$^{1,2,3,}$\thanks{Email: padovani@stsci.edu}, Raffaella Morganti$^{4}$, Joachim
Siebert$^5$, \newauthor Fausto Vagnetti$^3$, Andrea Cimatti$^6$
\\
$^1$ Space Telescope Science Institute, 3700 San 
Martin Drive, Baltimore MD. 21218, USA \\ 
$^2$ Affiliated to the Astrophysics Division, Space Science Department, 
European Space Agency \\
$^3$ Dipartimento di Fisica, II Universit\`a di Roma
``Tor Vergata'', Via della Ricerca Scientifica 1, I-00133 Roma, Italy \\
$^4$ Istituto di Radioastronomia, Via Gobetti 101, 
I-40129 Bologna, Italy \\
$^{5}$ Max-Planck-Institut f\"ur Extraterrestrische
Physik, Giessenbachstrasse,  D-85740 Garching bei M\"unchen, Germany\\
$^6$ Osservatorio Astrofisico di Arcetri,
Largo E. Fermi 5, I-50125 Firenze, Italy
}

\date{Accepted~~, Received~~}

\maketitle

\begin{abstract}

We present new \sax LECS, MECS and PDS observations of five lobe-dominated,
broad-line active galactic nuclei selected from the 2-Jy sample of southern
radio sources.  These include three radio quasars and two broad-line radio
galaxies.  ROSAT PSPC data, available for all the objects, are also used to
better constrain the spectral shape in the soft X-ray band.  The collected
data cover the energy range $0.1 - 10$ keV, reaching $\sim 50$ keV for one
source (Pictor A).  The main result from the spectral fits is that {\it all}
sources have a hard X-ray spectrum with energy index $\alpha_{\rm x} \sim
0.75$ in the $2 - 10$ keV range. This is at variance with the
situation at lower energies where these sources exhibit steeper spectra. 
Spectral breaks $\Delta \alpha_{\rm x} \sim 0.5$ at $1 - 2$ keV characterize 
in fact
the overall X-ray spectra of our objects.  The flat, high-energy slope is very
similar to that displayed by flat-spectrum/core-dominated quasars, which
suggests that the same emission mechanism (most likely inverse Compton)
produces the hard X-ray spectra in both classes.  Finally, a (weak) thermal
component is also present at low energies in the two broad-line radio galaxies
included in our study. 

\end{abstract}


\section{Introduction}

There is abundant evidence that strong anisotropies play a major role in the
observed characteristics of radio loud active galactic nuclei (AGN; see
Antonucci 1993 and Urry \& 
Padovani 1995 for a review). Radio jets are in fact known to be strongly
affected by relativistic beaming, while part of the optical emission in some
classes of objects is likely to be absorbed by a thick disk or
torus around the active nucleus. 

A unification of all high-power radio sources has been suggested (Barthel
1989; Urry \& Padovani 1995 and references therein) and according to this
scheme, the lobe-dominated, steep-spectrum radio quasars (SSRQ) and the
core-dominated, flat-spectrum radio quasars (FSRQ) are believed to be
increasingly aligned versions of Fanaroff-Riley type II (FR~II; Fanaroff \&
Riley 1974) radio galaxies. Within this scheme, broad-line (FWHM $\approxgt$
2000~\kms) radio galaxies (BLRG) have a still uncertain place.  They could
represent either objects intermediate between quasars and radio galaxies, 
(i.e., 
with the nucleus only partly obscured and the broad emission lines just 
becoming visible at the edge of the obscuring torus) or low-redshift, 
low-power equivalent of quasars. 

The scenario described above makes a number of predictions about the X-ray
emission of these radio-loud AGN. Moreover, the hard X-ray band, that is less
affected by absorption, is essential for a complete knowledge of the intrinsic
nature of these objects. 

Although the X-ray spectrum can be very complex, the presence of a nuclear, 
likely beamed X-ray component in quasars is quite well established in
particular for FSRQ and blazars (Wilkes \& Elvis 1987; Shastri \etal 1993;
Sambruna et al. 1994). There are mainly two arguments to support this: 1) the
tendency for radio loud AGN to have systematically flatter X-ray slopes than
radio quiet ones; 2) the fact that the soft X-ray slope decreases with core
dominance (Shastri \etal 1993) and increases with radio spectral index
(Fiore et al. 1998). Both these results are explained 
with the presence of a
radio-linked synchrotron self-Compton component of the X-ray emission that is
likely to be beamed. This component would be dominant in the FSRQ. In SSRQ, in
which ``blazar-like'', non-thermal emission is probably less important because
of the larger angle w.r.t. the line of sight, the ``UV bump'' would be
stronger (as effectively observed: e.g., Wills et al. 1995) and the steeper
soft X-ray component would represent its high-energy tail.

Although a nuclear X-ray component has been detected also in
radio galaxies, it appears to be much weaker than in radio quasars 
(consistent with the idea that
radio galaxies have an obscured nucleus).  For example, the X-ray spectrum of
Cygnus~A (Ueno \etal 1994) is consistent with a typical quasar spectrum
absorbed by a high column density of cold gas along the line of
sight. On the other hand, in the case of the broad-line radio galaxy 3C 390.3
(Inda \etal 1994), its hard X-ray spectrum can be described by a relatively
flat, unabsorbed power-law. This would suggest that BLRG might well be the
low-redshift counterpart of radio quasars and therefore aligned within
$\approxlt 40^{\circ}$ (as predicted by unified schemes: see e.g., Urry \&
Padovani 1995). 

{}From the above it is clear that a spectral X-ray study of lobe-dominated,
broad-line radio sources (including both SSRQ and BLRG), covering a large
X-ray band is necessary for a number of reasons. Namely: 1) to
study the hard X-ray properties of lobe-dominated, broad-line radio sources,
at present not well known; 2) to investigate if the difference in the soft
X-ray spectra of SSRQ and FSRQ apply also to the hard X-ray band. The
detection of a flatter component in SSRQ will be extremely important for our
understanding of the emission processes in this class of objects; 3) to
increase the number of BLRG for which the X-ray spectrum is known in detail in
order to disentangle the real nature of BLRG and investigate if the X-ray
spectra of BLRG and SSRQ are similar. 

In this paper we present \sax observations of five lobe-dominated, broad-line
radio sources, namely three SSRQ and two BLRG (we follow the commonly adopted
definition of lobe-dominated source, which implies a value of the core
dominance parameter $R < 1$). The sample is well 
defined (i.e., it is not a compilation of known hard X-ray sources)
and it is extracted from the 2-Jy sample of radio sources for which
a wealth of radio and optical information is available. The unique capability of
the \sax satellite (Boella \etal 1997a) of performing  simultaneous broad-band
X-ray ($0.1-200$ keV) studies is particularly well suited for a detailed
analysis of the X-ray energy spectrum of these sources. 

In \S~2 we present our sample, \S~3 discusses the observations and the data
analysis, while \S~4 describes the results of our spectral fits to the \sax
data. In \S~5 we also examine the ROSAT PSPC data of our sources to better
constrain the fits at low energies, in \S~6 we combine the analysis of the
\sax and ROSAT data while in \S~7 we briefly comment on the lack of iron lines
in our spectra. Finally, \S~8 discusses our results and \S~9 summarizes our
conclusions. Throughout this paper spectral indices are written $S_{\nu}
\propto \nu^{-\alpha}$.


\section{The Sample}

The lobe-dominated, broad-line objects studied in this paper belong 
to a complete subsample of the 2~Jy catalogue of radio sources (Wall \&
Peacock 1985). This subsample, defined by redshift $z< 0.7$ and declination
$\delta < 10^{\circ}$, includes 88 objects and is complete down to a flux
density level of 2 Jy at 2.7 GHz. Optical spectra are available for all the
sources together with accurate measurements of the 
[O~{III}]$\lambda$5007,
[O~{II}]$\lambda$3727 and H$\beta$ emission line fluxes 
(Tadhunter \etal 1993,
1998).  Estimates of the core dominance parameter $R$ [$R =
S_{core}/(S_{tot}-S_{core})$] have been derived from both arcsec-resolution
images and higher resolution data (Morganti et al. 1993, 1997). 
A study of the soft X-ray characteristics of the objects in the sample has
been carried out using the ROSAT All-Sky Survey and/or ROSAT PSPC pointed
observations (Siebert \etal 1996). For most of the objects, however, no
useful X-ray spectral information is available. 

The 2~Jy subsample described above contains 16 lobe-dominated, broad-line
objects (excluding compact steep-spectrum sources, whose relation to other
classes is still not clear). From those, we have selected the 10 sources with
estimated flux in the 0.1 -- 10 keV band larger than $2 \times 10^{-12}$ erg
cm$^{-2}$ s$^{-1}$ for an X-ray spectral study with the {\it BeppoSAX}
satellite\footnote{Note that, as expected in any flux-limited sample, the 10 
selected objects are $\sim 30$ times more luminous in the X-ray band than the
6 sources which did not make the X-ray flux cut. Our sample is then biased
towards the most X-ray luminous lobe-dominated, broad-line sources in the 2-Jy
sample.}. 
Here we present the results obtained for the 5 objects so far
observed in Cycle 1. The list of objects and their basic characteristics 
are given in Table 1, which presents the source name, position, redshift, 
optical magnitude $V$, 2.7 GHz radio flux, radio spectral index $\alpha_{\rm
r}$ (taken from Wall \& Peacock 1985), core dominance parameter $R$ at 2.3 GHz, Galactic \nh~and classification. 

\begin{table*}
\begin{center}
{\bf Table 1. Sample Properties}
\begin{tabular}{lrrrrrrrrr}
Name & RA(J2000) & Dec(J2000)&$z$ & V & F$_{\rm 
2.7GHz}$&$\alpha_{\rm 2.7-5~GHz}$ & R$_{\rm 2.3GHz}$ &
Galactic N$_{\rm H}$ & Class \\
     &           &           &   & &Jy       & & & $10^{20}$ cm$^{-2}$& \\
OF $-$109&04 07 48.4&$-$12 11 36&0.574 &14.9&2.35& 0.42& 0.57 & 3.81 & QSO \\
Pictor A &05 19 49.0&$-$45 46 46&0.035 &15.8&29.00& 1.07& 0.03 & 4.15 & BLRG \\
OM $-$161&11 39 10.6&$-$13 50 43&0.554 &16.2&2.80& 0.65& 0.16 & 3.59 & QSO \\
PHL 1657&21 37 44.1&$-$14 32 55&0.200 &15.5& 2.00& 0.63& 0.06 & 4.45 & QSO \\
PKS 2152$-$69&21 57 05.7&$-$69 41 23&0.027 &13.8&19.27& 0.71 & 0.03 & 2.50 & BLRG \\
\end{tabular}
\end{center}
\end{table*}

\section{Observations and Data Analysis}

A complete description of the \sax mission is given by Boella \etal (1997a). 
The relevant instruments for our observations are the coaligned Narrow Field
Instruments (NFI), which include one Low Energy Concentrator Spectrometer
(LECS; Parmar \etal 1997) sensitive in the 0.1 -- 10 keV band; three identical
Medium Energy Concentrator Spectrometers (MECS; Boella \etal 1997b), covering
the 1.5 -- 10 keV band; and the Phoswich Detector System (PDS; Frontera \etal
1997), coaligned with the LECS and the MECS.  The PDS instrument is made up of
four units, and was operated in collimator rocking mode, with a pair of units
pointing at the source and the other pair pointing at the background, the two
pairs switching on and off source every 96 seconds.  The net source spectra
have been obtained by subtracting the `off' to the `on' counts.  A journal of
the observations is given in Table 2.


The data analysis was based on the linearized, cleaned event files obtained
from the \sax Science Data Center (SDC) on-line archive (Giommi \& Fiore 1997)
and on the XIMAGE package (Giommi \etal 1991) upgraded to support the analysis
of \sax data. The data from the three MECS instruments were merged in one 
single event file by SDC. 
The LECS data above 4 keV were not used due to calibration
uncertainties in this band that have not been completely solved at this time
(Orr et al. 1998).
As recommended by the SDC, LECS data have been then fitted only in the $0.1-4$
keV range, while MECS data were fitted in the $1.8-10.5$ keV range. 

Spectra were accumulated for each observation using the SAXSELECT tool, with
8.5 and 4 arcmin extraction radii for the LECS and MECS respectively, which
include more than 90\% of the flux. The count rates given in Table 2 were
obtained using XIMAGE and refer to channels 10 to 950 for the LECS and 36 to
220 for the MECS. The \sax images were also checked for the presence of
serendipitous sources which could affect the data analysis. The ROSAT public
images of our sources were also inspected (see below). Most of our objects
have at least one source within the LECS extraction radius in the ROSAT PSPC
and/or MECS images but at a flux level which is at maximum 10\% of the target
(and in most cases below 3\%). Serendipitous sources in the field are then
unlikely to affect our results at a significant level. We also looked for
variability on timescales of 500 and 1,000 seconds for each observation, with
null results. 

The LECS and MECS background is low, although not uniformly distributed across
the  detectors, and rather stable. For this reason, and in particular for the
spectral analysis, it is better to evaluate the background from blank fields, 
rather than in annuli around the source. Background files accumulated from 
blank fields, obtained from the SDC public ftp site, were then used.

\begin{table*}
{\bf Table 2. Journal of observations}
\begin{tabular}{lrrrrr}
Name &LECS& LECS &MECS & MECS & Observing date\\
 &exposure (s)& count rate ($10^{-2}$ s$^{-1}$)&exposure (s)& count rate ($10^{-2}$ 
s$^{-1}$)& \\
OF $-$109&...~~~&...~~~~~&16,864& $~7.8 \pm 0.3$&22-SEP-1996\\
Pictor A &~5,271&$16.5\pm0.7$&14,798& $26.4 \pm 0.5$&12/13-OCT-1996\\
OM $-$161&13,782&$~3.3\pm0.2$&28,248& $ 5.7 \pm 0.2$&11/12-JAN-1997\\
PHL 1657&~4,704&$~9.3\pm0.6$&17,429& $18.0 \pm 0.4$&29-OCT-1996\\
PKS 2152$-$69& ...~~~& ...~~~~~& ~9,367& $15.1 \pm 0.5$&29-SEP-1996\\
\end{tabular}
\end{table*}

\section{Spectral Fits}

Spectral analysis was performed with the XSPEC 9.00 package, using the response
matrices released by SDC in early 1997. The spectra were rebinned such that 
each new bin contains at least 20 counts (using the command GRPPHA within 
FTOOLS). Various checks using some of the rebinning files provided by SDC 
have shown that our results are independent of the adopted rebinning within
the uncertainties.
The X-ray spectra of our sources are shown in Figure 1 (which includes both
\sax and ROSAT data: see Sect. \ref{sec:whole}). 

\subsection{LECS Data: Constraining N$_{\rm H}$}
\label{sec:lecs}
At first, we fitted the LECS data with a single power-law model with
Galactic and free absorption. 
The absorbing column was parameterized in terms of N$_{\rm H}$, the HI column
density, with heavier elements fixed at solar abundances. Cross sections were
taken from Morrison and McCammon (1983). For one set of fits \nh~was fixed at
the Galactic value, derived from Elvis, Lockman \& Wilkes (1989) for PHL 1657
and from the {\tt nh} program at HEASARC (based on Dickey \& Lockman
1990), for the remaining objects. \nh~was also set free to vary to check for
internal absorption and/or indications of a ``soft-excess.'' 


Our results are presented in Table 3, which gives the name of the source in
column (1), the energy index $\alpha_{\rm x}$ and reduced chi-squared and
number of degrees of freedom, $\chi^2_{\nu}$(dof), in columns (2)-(3) for the
fixed-\nh~fits; columns (4)-(6) give \nh, $\alpha_{\rm x}$ and
$\chi^2_{\nu}$(dof) for the free-\nh~fits.  Finally, in column (7) we report
the unabsorbed X-ray flux in the $0.1 - 4.0$ keV range (multiplied by a
normalization constant derived from the combined LECS plus MECS fits: see next
section).  The errors quoted on the fit parameters are the 90\% uncertainties
for one and two interesting parameters for Galactic and free \nh~respectively. 

Two results are immediately apparent from Table 3: the fitted energy indices
are flat (\ax~$< 1$); and the fitted \nh~values are consistent with the
Galactic ones. This is confirmed by an $F$-test which shows that the addition
of \nh~as a free parameter does not result in a significant improvement in the
$\chi^2$ values. We will then assume Galactic \nh~in the combined LECS 
and MECS fits. For the two objects without LECS data this assumption is also 
justified by the fact that the fit to the MECS data is not strongly dependent on 
\nh. 

The spectrum of PHL 1657 is more complicated than a simple power-law: the 
residuals show a clear excess at $E \approxlt 0.7$ keV. Indeed a broken 
power-law model significantly improves the fit (see below).
Weaker ``soft-excesses'' cannot be excluded in the two other sources (see 
below) so the fluxes given in Table 3, based on a single power-law fit to the 
data, are almost certainly underestimated.


\begin{table*}
{\bf Table 3. LECS spectral fits}
\begin{tabular}{@{}lcccccr}
      & ~~~~~~~~~ Galactic & N$_{\rm H}$ & ~~~~~~~~~ Free&N$_{\rm H}$ & \\
Name &$\alpha_{\rm x}^a$& $\chi^2_{\nu}$(dof)&N$_{\rm H}^b$&$\alpha_{\rm x}^b$& 
$\chi^2_{\nu}$(dof) &Flux (0.1 -- 4.0 keV)$^c$\\
     &   &   & $10^{20}$ cm$^{-2}$& & & $10^{-12}$ erg cm$^{-2}$ s$^{-1}$\\

Pictor A&0.69$^{+0.14}_{-0.14}$&0.78(30)&$2.75^{+2.79}_{-1.55}$&
0.60$^{+0.23}_{-0.22}$&0.75(29) & $17.6 \pm 0.7 $\\
OM $-$161&0.77$^{+0.20}_{-0.22}$&0.77(24)&$3.70^{+7.80}_{-2.46}$&
0.77$^{+0.40}_{-0.36}$&0.80(23) & $ 3.8 \pm 0.2 $\\ 
PHL 1657$^d$&0.83$^{+0.23}_{-0.24}$&1.01(16)&$2.75^{+3.48}_{-1.98}$&
0.69$^{+0.38}_{-0.36}$&0.97(15) & $13.5 \pm 0.9 $\\ 
\end{tabular}

\noindent
$^a$ Quoted errors correspond to 90\% uncertainties for one interesting
parameter. \\
$^b$ Quoted errors correspond to 90\% uncertainties for two interesting
parameters. \\
$^c$ Unabsorbed flux. 1 $\sigma$ statistical errors for best-fit model (model
uncertainties not included). LECS fluxes have been multiplied by a
normalization constant derived from the combined LECS and MECS fits ($\sim
0.6^{-1} - 0.8^{-1}$: see text). \\
$^d$ An $F$-test shows that a broken power-law model improves the fit at the
$98.1\%$ level. Best-fit parameters are (for Galactic N$_{\rm H}$):
$\alpha_{\rm S} = 1.2$, $\alpha_{\rm H} = 0.3$, $E_{\rm break} = 1.6$ keV,
$\chi^2_{\nu}$(dof) = 0.84(14). \\
 
\end{table*}

\subsection{LECS and MECS Data}
\label{sec:lecsmecs}

Our results from the jointly fitted LECS and MECS data assuming a single
power-law model with Galactic absorption are presented in Table 4, which gives
the name of the source in column (1), $\alpha_{\rm x}$ and $\chi^2_{\nu}$(dof)
in columns (2)-(3), and the unabsorbed X-ray flux in the $2 - 10$ keV range in
column (4).  The errors quoted on \ax~are 90\% uncertainties. 

Due to the uncertainties in the calibration of the LECS
instrument, the LECS/MECS normalization has been let free to vary. The
resulting values, in the 0.6 -- 0.8 range, are consistent with the expected
one (Giommi \& Fiore, private communication). 

\begin{figure}
\centerline{\psfig{figure=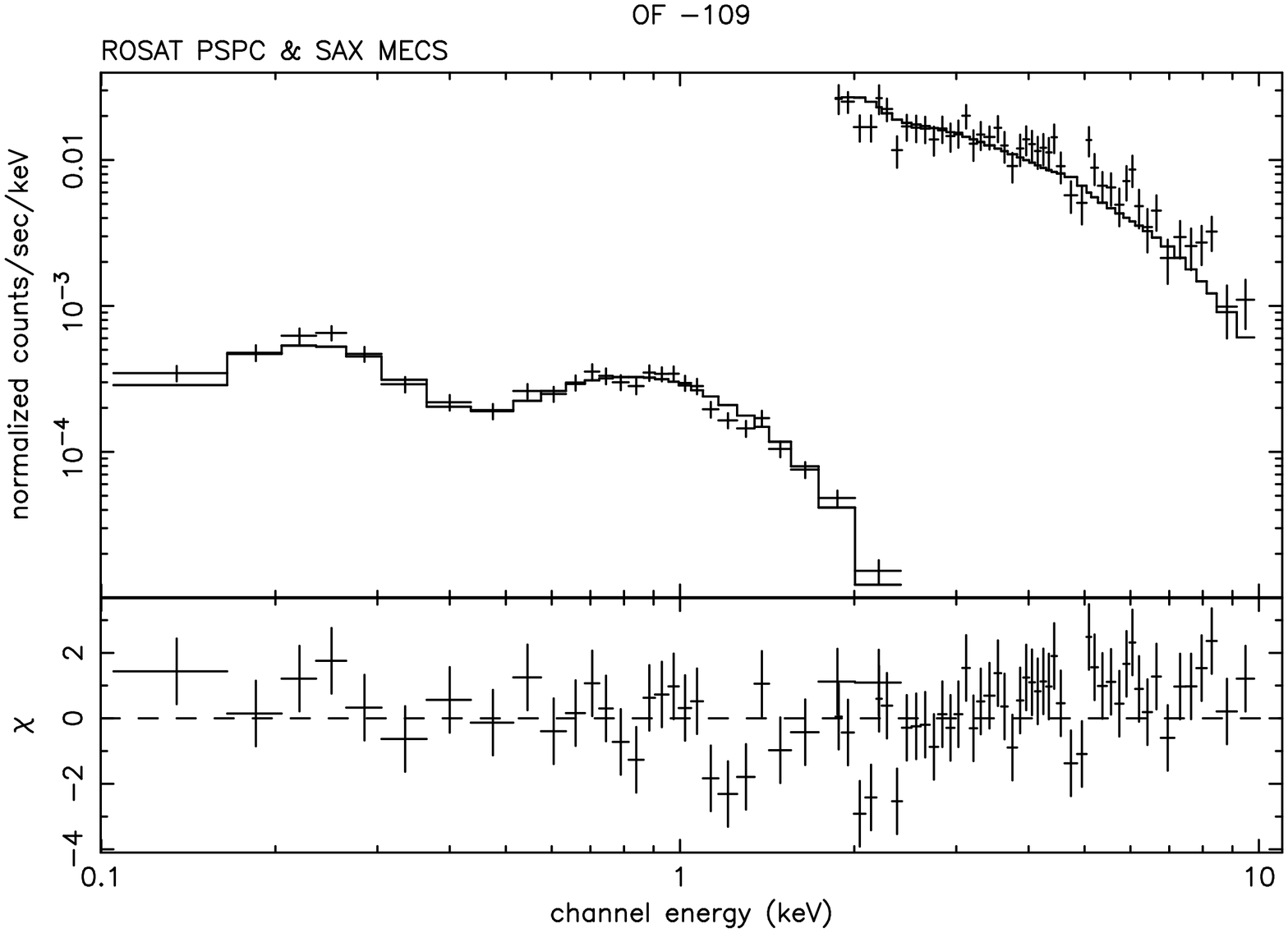,width=9cm,angle=-90}}
\centerline{\psfig{figure=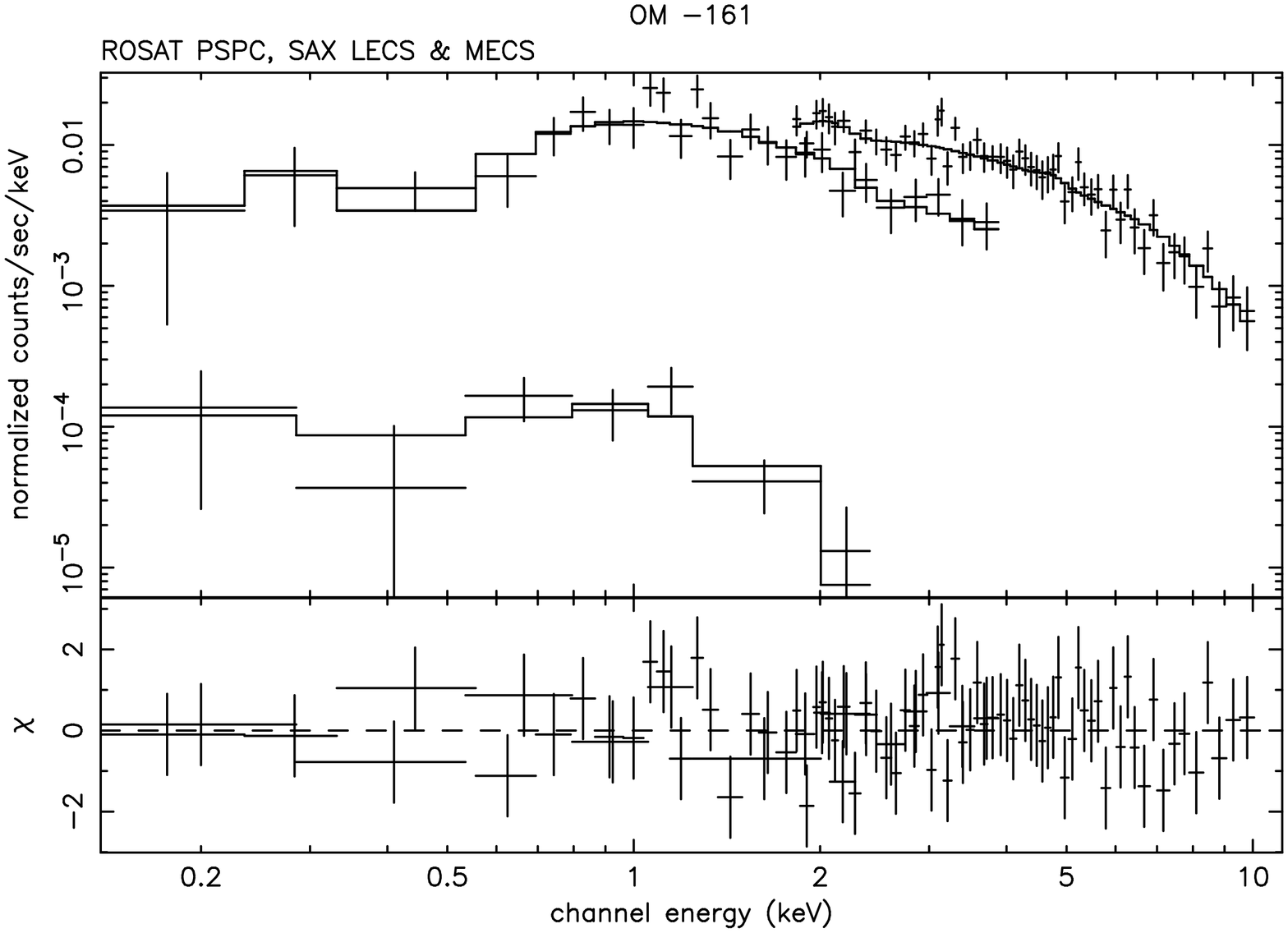,width=9cm,angle=-90}}
\centerline{\psfig{figure=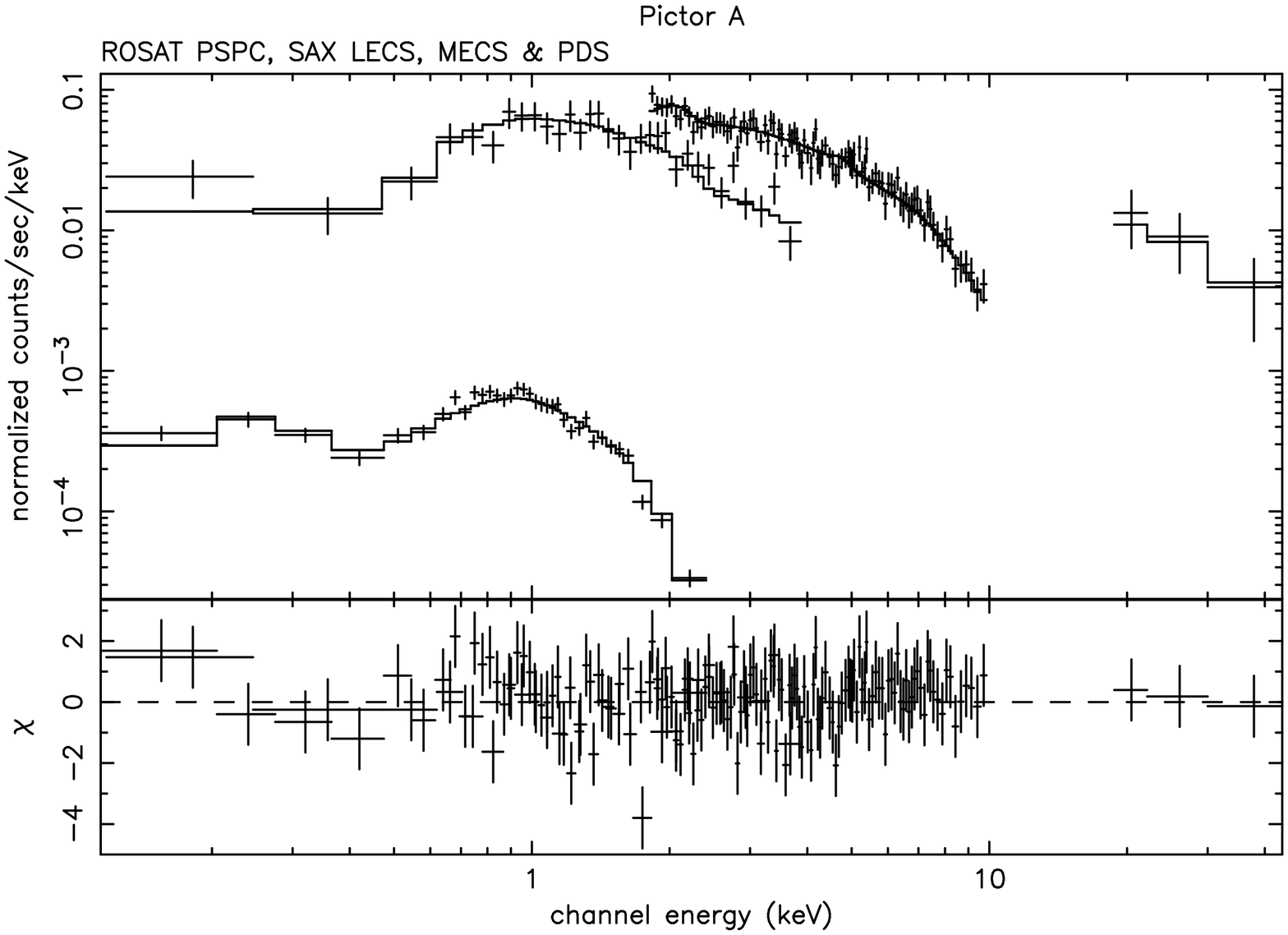,width=9cm,angle=-90}}
\end{figure}
\begin{figure}
\centerline{\psfig{figure=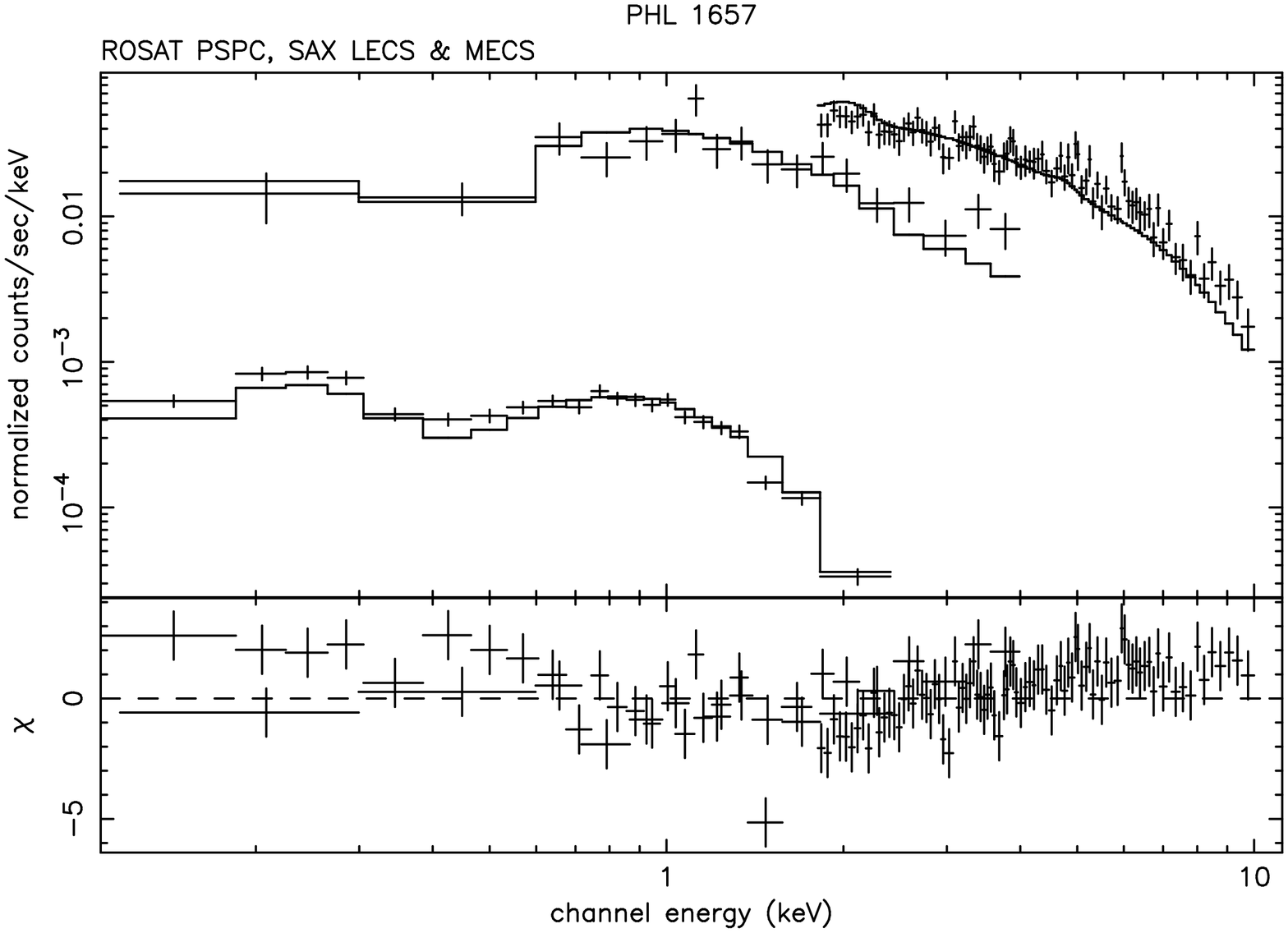,width=9cm,angle=-90}}
\centerline{\psfig{figure=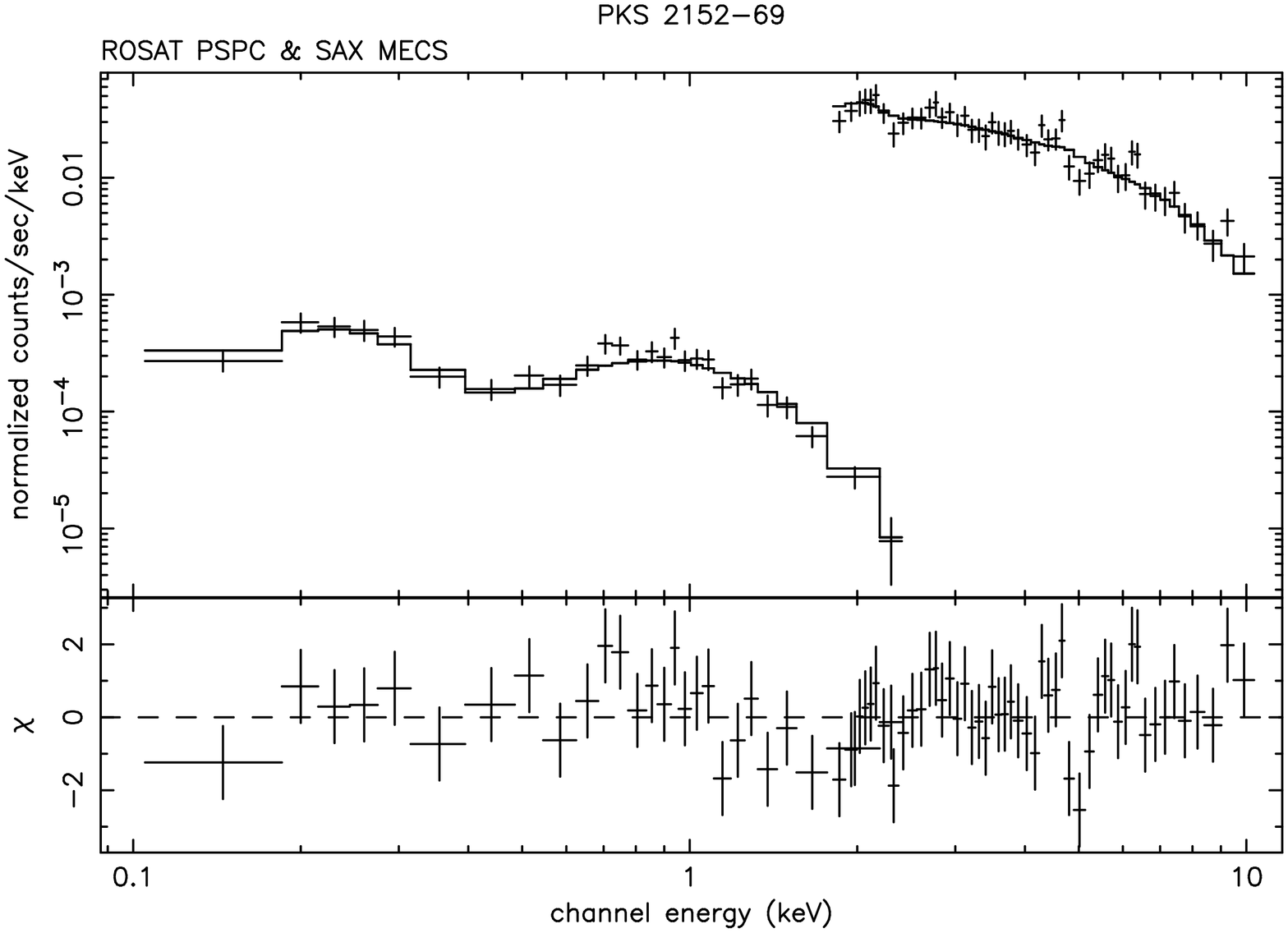,width=9cm,angle=-90}}
\caption{Spectra of the combined \sax and ROSAT PSPC data for our
sources. The data are fitted with a single power-law model with Galactic
absorption (free absorption for PKS 2152$-$69). Note that the ROSAT counts
[lower spectra] have been normalized by the PSPC geometric area of 1,141
cm$^2$ and should be read as ``normalized counts/sec/keV/cm$^2$.''
}  

\end{figure}

The striking result is that {\it all} the
sources have relatively flat X-ray energy indices. The mean value is $\langle 
\alpha_{\rm x} \rangle = 0.75\pm0.02$ (here and in the following we give the
standard deviation of the
mean). This implies that the spectra are still raising in a $\nu - f_{\nu}$
plot, and therefore that the peak of the X-ray emission in the \sax band is at
$E > 10$ keV. 
Table 4 also reports (in the footnotes) the fits to the MECS data for the
three sources with both LECS and MECS observations. The energy indices in the
$1.8 - 10.5$ keV range have a mean value $\langle \alpha_{\rm x} \rangle =
0.74\pm0.03$, basically the same as in the whole $0.1 - 10.5$ keV band. 


As mentioned in 
the Introduction, the spectra of the class of sources under study are
generally steep at lower X-ray energies (and  
there is indeed strong evidence for a steeper X-ray component in the LECS data
of PHL 1657). 
We then tried to fit a broken power-law model to our data. A
significant improvement in the fit (96.5\% level) was obtained only for PHL
1657, whose residuals again showed a clear excess at $E \approxlt 0.7$ keV.
The best-fit parameters are $\alpha_{\rm S} = 1.3$, $\alpha_{\rm H} =
0.76\pm0.12$, and $E_{\rm break} = 0.9$ keV. 

The fact that the other four sources show no significant evidence for a
concave spectrum needs to be investigated with more data 
at soft X-ray energies. Hence the need
to resort to ROSAT PSPC data (see Sect. \ref{sec:pspc}). 


\subsection{The PDS Detection of Pictor A}
\label{sec:pds}
Only the brightest source of our sample, Pictor A, has been detected by 
the
PDS instrument (up to $\sim 50$ keV; see Fig. 1) despite the relative short
exposure time (6.8 ks). The count rate is $0.18\pm0.05$ ct/s, that is the
significance of the detection is about 3.6 $\sigma$. Given the relatively
small statistics, it is hard to constrain the high energy ($E \ga 10$ keV)
spectrum of Pictor A. A parametrization of the MECS and PDS data with a single
power-law model gives a best fit value of $\alpha_{\rm x}$ perfectly
consistent with that derived from MECS data only. A broken power-law model,
with the soft energy index fixed to the value obtained from the MECS data (see
Table 4), gives no significant improvement in the fit (and the hard energy
index is consistent with the soft one). Therefore, the PDS data appear to lie
on the extrapolation of the lower energy data. There might be a slight excess
in the residuals above 10 keV but as described above this is not significant
and does not warrant more complicated models.

\section{ROSAT PSPC Data}
\label{sec:pspc}
All our objects were found to have ROSAT PSPC
data: namely, OF $-109$, Pictor A, PHL 1657 and PKS 2152$-$69 were all
targets of ROSAT observations, while data for OM $-161$ were extracted from
the ROSAT All-Sky Survey.

\subsection{Data Analysis}

In the analysis of the pointed PSPC observations, we first determined
the centroid X-ray position by fitting a two-dimensional Gaussian to the X-ray
image. Source counts were then extracted from a circular region
with 3 arcmin radius around the centroid source position. The local
background was determined from an annulus with inner radius 5 arcmin and
outer radius 8 arcmin. If any X-ray sources were detected in the background
region, they were first subtracted from the data.

The source counts from OM $-161$ were extracted from a circular region with
radius 5 arcmin from the All-Sky Survey data. The larger extraction radius
compared to the pointed PSPC observations accounts for the larger point
spread function in the Survey. The local background was determined from two
source-free regions with radius 5 arcmin, displaced from the source position
along the scanning direction of the satellite during the All-Sky Survey.

After background subtraction, the data were vignetting and dead time corrected
and finally binned into pulse height channels. Only channels 12-240 were used
in the spectral analysis, due to existing calibration uncertainties at lower
energies. The pulse height spectra were rebinned to achieve a constant
signal-to-noise ratio in each spectral bin, which ranged from 3 to 6,
depending on the total number of photons.

\begin{table}
\begin{center}
{\bf Table 4. LECS and MECS spectral fits}
\begin{tabular}{lclr}
Name & $\alpha_{\rm x}^a$&
$\chi^2_{\nu}$(dof)& Flux (2 -- 10 keV)$^b$ \\
     & &  & $10^{-12}$ erg cm$^{-2}$ s$^{-1}$ \\
OF $-$109$^c$&
0.81$^{+0.14}_{-0.14}$&1.17(46) & $ 3.7 \pm 0.1 $ \\
Pictor A$^d$& 
0.68$^{+0.06}_{-0.07}$&0.78(128)&$13.8 \pm 0.3 $ \\ 
OM $-$161$^e$& 
0.77$^{+0.11}_{-0.11}$&0.81(80)& $2.5 \pm 0.1$ \\
PHL 1657$^{f,g}$& 
0.78$^{+0.08}_{-0.08}$&0.82(105)& $8.6 \pm 0.2$ \\
PKS 2152$-$69$^c$& 
0.70$^{+0.12}_{-0.12}$&1.04(48)& $7.5 \pm 0.2$ \\
\end{tabular}
\end{center}
$^a$ Assuming Galactic N$_{\rm H}$. \\
$^b$ Unabsorbed flux. 1 $\sigma$ statistical errors for best-fit model (model
uncertainties not included). \\
$^c$ Only MECS data available. \\
$^d$ MECS only fit: $\alpha_{\rm x} = 0.67^{+0.07}_{-0.07}$, $\chi^2_{\nu}$(dof) 
= 0.78(97) \\
$^e$ MECS only fit: $\alpha_{\rm x} = 0.77^{+0.13}_{-0.12}$, $\chi^2_{\nu}$(dof) 
= 0.84(55) \\
$^f$ MECS only fit: $\alpha_{\rm x} = 0.77^{+0.08}_{-0.08}$, $\chi^2_{\nu}$(dof) 
= 0.79(88) \\
$^g$ An $F$-test shows that a broken power-law model improves the fit at the
$96.5\%$ level. Best-fit parameters are: $\alpha_{\rm S} = 1.3$, $\alpha_{\rm
H} = 0.76\pm0.12$, $E_{\rm break} = 0.9$ keV, $\chi^2_{\nu}$(dof)= 0.81(103). \\
\end{table}

\subsection{Spectral Fits}

As for the LECS data, we fitted the ROSAT PSPC data with a single power-law
model with Galactic and free absorption. Our results are presented in Table 5,
which gives the name of the source in column (1), the ROSAT observation 
request (ROR) number in column (2), $\alpha_{\rm x}$ and
$\chi^2_{\nu}$(dof) in columns (3)-(4) for the fixed-\nh~fits; columns
(5)-(7) give \nh, $\alpha_{\rm x}$ and $\chi^2_{\nu}$(dof) for the free-\nh~
fits. Finally, in column (8) we report the unabsorbed X-ray flux in the $0.1 -
2.4$ keV range. The errors quoted on the fit parameters are the 90\%
uncertainties for one and two interesting parameters for Galactic and free
\nh~respectively. 

The main results of the ROSAT PSPC fits are the following: 1. the fitted
energy indices are steeper than the MECS (plus LECS) ones (with the exception
of OM $-161$ for which the ROSAT \ax~has large uncertainties); 2. there is no
evidence for intervening absorption above the Galactic value in our sources,
with the exception of PKS 2152$-$69, for which the $F$-test shows that the
addition of \nh~as a free parameter results in a significant improvement
(98.6\% level) in the goodness of the fit (the fitted \nh~is about 50\% higher
than the Galactic value); 3. the single power-law fit is not great, although
still acceptable, 
for Pictor A ($P_{\chi^2} \sim 5\%$) and PHL 1657 ($P_{\chi^2} \sim 7\%$).

The mean difference between the ROSAT PSPC and MECS (plus LECS) energy indices 
(excluding OM $-161$) is $0.44\pm0.11$, clearly indicative of a flattening
at high energies, with the emergence of a hard component.

\begin{table*}
\begin{center}
{\bf Table 5. ROSAT PSPC spectral fits}
\begin{tabular}{lcclcclr}
      & & Galactic&N$_{\rm H}$ &  Free&N$_{\rm H}$ & \cr
Name & ROR \#&$\alpha_{\rm x}^a$& $\chi^2_{\nu}$(dof)&N$_{\rm H}^b$&$\alpha_{\rm x}^b$& 
$\chi^2_{\nu}$(dof) &Flux (0.1 -- 2.4 keV)$^c$\\
     &  &  &   & $10^{20}$ cm$^{-2}$& & & $10^{-12}$ erg cm$^{-2}$ s$^{-1}$\\
OF $-$109&701072&1.30$^{+0.07}_{-0.06}$&1.04(26)&$3.65^{+0.77}_{-0.72}$&
1.26$^{+0.23}_{-0.23}$&1.08(25) & $11.9\pm0.3$ \\
Pictor A&700057&0.80$^{+0.05}_{-0.05}$&1.46(30)&$4.83^{+0.68}_{-0.65}$&
0.96$^{+0.16}_{-0.17}$&1.34(29)$^d$ & $17.7\pm0.3$ \\
OM $-$161&...&0.80$^{+0.55}_{-0.80}$&0.65(5)&$6.2^{~~~~}_{~~~~}$&
1.3$^{~~~~}_{~~~~}$& 0.73(4)$^e$ & $3.9\pm0.8$ \\
PHL 1657&701542&1.42$^{+0.06}_{-0.05}$&1.49(20)&$4.86^{+0.75}_{-0.72}$&
1.53$^{+0.22}_{-0.21}$&1.50(19) & $23.7\pm0.5$ \\ 
PKS 2152$-$69&701154&0.86$^{+0.09}_{-0.09}$&1.02(25)&$3.72^{+1.22}_{-1.10}$&
1.22$^{+0.36}_{-0.35}$&0.83(24)$^f$ & $6.4\pm0.2$ \\
\end{tabular}
\end{center}

$^a$ Quoted errors correspond to 90\% uncertainties for one interesting
parameter. \\
$^b$ Quoted errors correspond to 90\% uncertainties for two interesting
parameters. \\
$^c$ Unabsorbed flux. 1 $\sigma$ statistical errors for best-fit model (model
uncertainties not included). \\
$^d$ The reduction in the $\chi^2$ value obtained with the free N$_{\rm H}$ 
fit is significant at the $93.7\%$ level according to the $F$-test. \\
$^e$ The errors on the best-fit parameters are essentially undetermined due to
the low photon statistics. \\
$^f$ The reduction in the $\chi^2$ value obtained with the free N$_{\rm H}$ 
fit is significant at the $98.6\%$ level according to the $F$-test. 
\end{table*}

\subsection{A Thermal Component in the X-ray Spectra of Pictor A and PKS
2152$-$69} 

Pictor A and PKS 2152$-$69 are relatively nearby objects ($z \le 0.035$).  The
ROSAT PSPC images show evidence for an extended component on a scale
$\approxgt 50"$ for PKS 2152$-$69 and $\approxgt 70"$ for Pictor A, which
correspond to about 40 and 70 kpc respectively.  
PKS~2152-69 is also seen extended from ROSAT HRI data (Fosbury, 
private communication).
Early-type galaxies are known to have diffuse emission from hot gas on these
scales (Forman, Jones \& Tucker 1985), so we added a thermal component
(Raymond \& Smith 1977) to the power-law model, assuming solar abundances (our
results are only weakly dependent on the adopted abundances).  
Our results are reported in Table 6 which gives the name of the source in
column (1), \ax~in column (2), the gas temperature (in keV) in column (3),
$\chi^2_{\nu}$(dof) in column (4), the ratio between the thermal and
non-thermal components in the $0.1 - 2.4$ keV range in column (5), and finally the
$F$-test probability in column (6). The errors quoted on the fit parameters
are the 90\% uncertainties for two interesting parameters. Galactic \nh~as
been assumed (see above). 
In both cases this addition results in a significantly improved fit ($> 99.9\%$
level) over a single power-law model.  (Note that a Raymond-Smith model by
itself gives extremely poor fits to the data.) With the addition of a thermal
component the need for absorption above the Galactic value, which was indicated
for PKS 2152$-$69 and suggested for Pictor A vanishes; free \nh~fits now do not
result in a significant improvement in the goodness of the fits.

\begin{table*}
\begin{center}
{\bf Table 6. ROSAT PSPC spectral fits: thermal and non-thermal 
components}
\begin{tabular}{lcccccc}
Name &$\alpha_{\rm x}^a$& $kT^a$&$\chi^2_{\nu}$(dof)& RS/power-law$^b$ & 
P($F$-test)$^c$\\
     &  & keV & &\\
Pictor A&0.79$^{+0.08}_{-0.10}$&0.55$^{+0.40}_{-0.30}$&1.13(28)&0.05&99.97\% \\
PKS 2152$-$69&0.91$^{+0.19}_{-0.19}$&0.59$^{+0.50}_{-0.37}$&0.73(23)&0.10&99.97\% \\
\end{tabular}
\end{center}

$^a$ Quoted errors correspond to 90\% uncertainties for two interesting
parameters. Galactic N$_{\rm H}$ assumed. \\
$^b$ Flux ratio of the Raymond-Smith and power-law components in the $0.1-2.4$
keV band. \\
$^c$ Probability that the decrease in $\chi^2$ due to the addition of the 
thermal component to the power-law model is significant.\\

\end{table*}

As a check of our results we also fitted the spectra of OF $-109$ and PHL 1657
with a power-law plus thermal component. No need for an extra component
was found, which is consistent with the fact that these two sources are at
higher redshift (i.e., the putative thermal component is completely swamped by
the stronger non-thermal emission). 

It is interesting to note that the dominant component in the X-ray emission of
our two nearest sources is definitely non-thermal, but nevertheless the data
indicate a $5 - 10\%$ contribution from thermal emission. This is confirmed by
an analysis of the PSPC images. The relevance of extended emission was in fact
estimated by subtracting the point spread function (PSF) from the radial
profile of the sources. For Pictor A, the fraction of photons above the PSF is
4\%, while for PKS 2152$-$69 is 13\% (for the two other sources with PSPC
pointed data these fractions are less than 1\%, as expected). Given the
statistical and systematic uncertainties in the PSF (due to residual wobble
motion, attitude uncertainties, etc.) these fractions agree very well with the
results from the spectral decomposition (5 and 10\% respectively).

The gas temperatures we find ($< 1$ keV) are very reasonable for gas 
associated with an
elliptical galaxy (e.g., Forman et al. 1985). To check that the observed
luminosities are also physically plausible, we performed the following test.
There is a well-known strong correlation between X-ray luminosity and absolute
blue magnitude for elliptical galaxies (e.g., Forman et al. 1985, 1994).
Integrated blue magnitudes for Pictor A and PKS 2152$-$69, obtained from NED,
imply $M_{\rm B} \simeq -20.7$ and $-22$ respectively. The $0.5 - 4.5$ keV
luminosities for the thermal components of the two sources are $L_{0.5-4.5} 
\simeq
4 \times 10^{42}$ erg s$^{-1}$ for Pictor A, with a rather large 90\% error
range ($10^{42} - 10^{43}$) while for PKS 2152$-$69 we get $L_{0.5-4.5} 
\simeq 2
\times 10^{42}$ erg s$^{-1}$ (90\% error range: $3 \times 10^{41} - 8 \times
10^{42}$). These numbers, compared against Fig. 4 of Forman et al. (1994),
show that while the X-ray power in the thermal component of PKS 2152$-$69 is
not unusual for its optical power, that of Pictor A is about an order of
magnitude larger than the {\it maximum} values of elliptical galaxies of the
same absolute magnitude. It then seems that the intrinsic power of the thermal
component is too large to be associated with the galaxy.

As the relatively low gas temperature inferred from the data is also typical
of small groups, one could speculate that most of the thermal emission in
Pictor A is associated with a group associated with this source. An inspection
of Fig. 8 of Ponman et al. (1996), which reports the X-ray luminosity --
temperature relation for Hickson's groups, shows that, within the rather large
errors, Pictor A might fall in the correct portion of the plot, although the
best fit values ($L_{\rm x} \simeq 4 \times 10^{42}$ erg s$^{-1}$, $kT = 0.55$
 keV),
would put it above the observed correlation. However, the richness of the
environment of this source is very low (Zirbel 1997), inconsistent even with a
small group. It might therefore be speculated that Pictor A is another example
for a so-called fossil group (Ponman et al. 1994), i.e. a single elliptical
galaxy that is considered to be the result of a merging process of a compact
group. This merging is believed not to affect the X-ray halo of the group
(Ponman \& Bertram 1993) and the galaxies formed in this way will still show
the extended thermal emission component of the intra-group gas although
they appear isolated.

In summary, while the thermal component in PKS 2152$-$69 is consistent with
emission from a hot corona around the galaxy, in the case of Pictor A the
intrinsic power of this component is too high. However, the luminosity of the
thermal component is consistent with that of a compact group of galaxies.
Since Pictor A appears to be isolated, we might have another example of a
fossil group.

We found no physically meaningful evidence for the presence of a thermal
component in the LECS spectra of the three sources for which we have the
relevant data. 


\section{ROSAT and \sax Data: The Whole Picture}
\label{sec:whole}
The last step is to put \sax and ROSAT PSPC data together to better
constrain the shape of the X-ray spectra, especially at low energies. 
Two of
our sources, in fact, have no LECS data, while for the remaining three only
less than 10 LECS bins are available below 1 keV. The ROSAT effective area is
larger than the LECS effective area in the range of overlap, providing more
leverage in the soft X-rays. 

As before, we left free the LECS/MECS normalization and, as
before, the fitted values are consistent with the expected ones. We also left
free the PSPC/MECS normalization, to allow for any X-ray variability, which is
seen in at least some lobe-dominated broad-line sources (e.g., 390.3: Leighly
et al.  1997; 3C 382: Barr \& Giommi 1992). However, the PSPC/MECS
normalization was, on average, around 1, with a maximum excursion of $30 -
40\%$. No strong variability is then present between the ROSAT and \sax data.

As it turned out, in all cases for which we had enough statistics at low
energies (i.e., excluding OM $-161$) a broken power-law model resulted in a
significantly improved fit ($\ge 99.9\%$ level) over a single power-law model
over the whole 0.1 -- 10.5 keV range. Our results are reported in Table 7 which
gives the name of the source in column (1), $\alpha_{\rm S}$, $\alpha_{\rm
H}$, and $E_{\rm break}$ in columns (2)-(4), $\chi^2_{\nu}$(dof) in column (5)
and finally the $F$-test probability in column (6). The errors quoted on the
fit parameters are the 90\% uncertainties for three interesting parameters.
Based on the LECS and ROSAT PSPC results, Galactic \nh~as been assumed for
all sources apart from PKS 2152$-$69. The combined data with the best fit 
single power-law model (to show the spectral concavity) are shown in Figure 1.

\begin{table*}
\begin{center}
{\bf Table 7. ROSAT PSPC, BeppoSAX LECS and MECS spectral fits}
\begin{tabular}{lccclr}
Name &$\alpha_{\rm S}^a$& $\alpha_{\rm H}^a$& $E_{\rm 
break}^a$&$\chi^2_{\nu}$(dof)& P($F$-test)$^b$\\
 &  &  & keV & & \\
OF $-$109&1.33$^{+0.11}_{-0.10}$&0.80$^{+0.20}_{-0.21}$&$1.32^{+1.39}_{-0.37}$
&1.06(71)& $>$99.99\% \\
Pictor A&0.79$^{+0.07}_{-0.07}$&0.56$^{+0.21}_{-0.38}$&$3.92^{+4.88}_{-3.10}$
&0.91(157)& 99.90\% \\
OM $-$161&...&0.77$^{+0.11}_{-0.11}$&...&0.79(86)&... \\
PHL 1657&1.42$^{+0.09}_{-0.09}$&0.75$^{+0.12}_{-0.12}$&$1.45^{+0.50}_{-0.29}$
&0.96(124)& $>$99.99\% \\
PKS 2152$-$69$^c$&1.23$^{+0.57}_{-0.41}$&0.70$^{+0.18}_{-0.18}$&$1.69^{+2.01}_{-0.75}$
&0.98(71)& $>$99.99\% \\
\end{tabular}
\end{center}

$^a$ Quoted errors correspond to 90\% uncertainties for three interesting 
parameters for all sources apart from OM $-161$ (one interesting parameter).
Galactic N$_{\rm H}$ assumed for all sources apart from PKS 2152$-$69. \\
$^b$ Probability that the decrease in $\chi^2$ due to the addition of two 
parameters (from a single power-law fit to a broken power-law fit) 
is significant.\\
$^c$ Fit with free $N_{\rm H} = 3.8^{+1.0}_{-0.9} \times 10^{20}$ cm$^{-2}$.

\end{table*}


As can be seen from the Table, the model parameters are extremely well
determined. Not surprisingly, the $\alpha_{\rm S}$ values are very similar to
the ROSAT PSPC energy indices, while the  $\alpha_{\rm H}$ values are
basically the same as the MECS (plus LECS) energy indices. The spectra are
obviously concave, with $\langle \alpha_{\rm S} - \alpha_{\rm H} \rangle = 0.49 \pm
0.09$ and energy breaks around 1.5 keV (Pictor A has a break at about 4 keV 
but with a large error due to the relatively small difference between the soft 
and the hard spectral indices). The fact that the breaks fall at relatively 
low energies explain why the energy indices derived from the LECS fits are 
basically the same as those obtained from the combined LECS and MECS fits.

The addition of a thermal component in Pictor A and PKS 2152$-$69 improves
significantly the fit ($> 99.8\%$ level) even in the case of a broken
power-law model. As for the single power-law plus thermal component, free
\nh~fits do not result in a significant improvement in the goodness of the
fits, so Galactic \nh~is assumed. Best-fit parameters are $\alpha_{\rm S} =
0.77$, $\alpha_{\rm H} = 0.66$, $E_{\rm break} = 2.1$ keV, $kT = 0.39$ keV, for
Pictor A, and $\alpha_{\rm S} = 0.92$, $\alpha_{\rm H} = 0.68$, $E_{\rm break}
= 1.4$ keV, $kT = 0.57$ keV, for PKS 2152$-$69. These are perfectly consistent
with those obtained from the broken power-law fit to the \sax and ROSAT PSPC
data and with the temperatures derived from the ROSAT PSPC data, but the
uncertainties are now poorly determined because of the relatively large number
of parameters. 

\section{Iron Lines?}

A number of AGN exhibit in their X-ray spectra iron K$\alpha$ lines which are
characteristic of relativistic effects in an accretion disk surrounding a 
central black hole (e.g., Nandra \& Pounds 1994). It appears that radio-loud 
AGN have weaker iron lines than radio-quiet ones, although some 
low-luminosity, 
radio-loud sources are known to have strong iron lines (Nandra et al. 1997). 

We searched for Fe K$\alpha$ emission in our MECS spectra: none was found. The
90\% upper limits on the equivalent width (in the source rest frame) of an
unresolved iron line 
($\sigma = 0$) at energy 6.4 keV are the following: OF $-109$: 380 eV; Pictor
A: 150 eV; OM $-161$: 300 eV; PHL 1657: 170 eV; PKS 2152$-$69: 400 eV. Note
that for any broader line the limits are correspondingly higher. Our result
for Pictor A is consistent with the upper limit of 100 eV given by Eracleous \&
Halpern (1998) based on a $\sim 65$ ks {\it ASCA} observation, and our limit 
for 
PHL 1657 agrees with the upper limit of 140 eV given by Williams et al. (1992)
based on a 15 ks {\it Ginga} observation. 

We note that the upper limits on iron lines in our sources are not very
stringent and still consistent with values found in other lobe-dominated
broad-lined sources (e.g., Inda et al. 1994; Lawson \& Turner 1997; Wo\'zniak
et al. 1998).

\section{Discussion}

We have presented the first systematic, hard X-ray study of a well-defined
sample of lobe-dominated, broad-line AGN. Our main result is that this
class of objects has a hard X-ray spectrum with \ax~$\sim 0.75$
at $E \approxgt 1 - 2$ keV. In addition, we also detect a thermal emission
component present at low energies in the spectra of two BLRGs, but we find
that this component contributes only $\approxlt 10\%$ of the total flux.

Hard X-ray emission is also present in core-dominated radio-loud quasars, and
the detection of similarly flat spectra in our sample of lobe-dominated AGN has
important implications for our understanding of the relation between the two
classes. 

In order to make a more quantitative comparison between the hard X-ray spectra
of the two classes we searched the literature for a study of
core-dominated/flat-spectrum radio quasars similar to ours, i.e., based on a
well-defined, homogeneous sample.  Surprisingly, we found none. We then
decided to collect all the information we could find on the $2 - 10$ keV
spectra of FSRQ, excluding objects with large ($> 0.5$) uncertainties on the
X-ray spectral index. The data come from {\it Ginga} and EXOSAT/ME
observations published in Makino (1989), Ohashi et al. (1989, 1992), Lawson et
al. (1992), Saxton et al. (1993), Williams et al. (1992), Sambruna et
al. (1994) and Lawson \& Turner (1997; $\alpha_{\rm x}$ in the $2 - 18$ keV
range). The resulting sample includes 15 objects and is characterized by
$\langle \alpha_{\rm x} \rangle = 0.70\pm0.06$, perfectly consistent
with our results.

We stress again that the sample of FSRQ is heterogeneous whereas our sample,
although relatively small, is well defined and has very well determined
spectral indices. However, within the limits introduced by the biases likely
present in the FSRQ sample, lobe-dominated and core-dominated broad-line AGN
appear to have practically identical hard X-ray spectra in the $2-10/18$ keV
region.

The objects in our sample are lobe dominated and therefore we 
investigated if inverse Compton scattering of cosmic microwave background
photons into the X-ray band by relativistic electrons in the diffuse radio
lobes could be responsible for the observed X-ray emission (see e.g., Harris
\& Grindlay 1979; Feigelson et al. 1995). We find that this is not the case
and the derived X-ray emission is
almost two orders of magnitude lower than observed.  
This is further supported by the fact that no strong resolved components have
been found for our objects by ROSAT.  The hotspot in Pictor A is known to have
associated X-ray emission but this is very weak and indeed only marginally
detected by Einstein (R\"oser \& Meisenheimer 1987; Perley et al.  1997). 
 
The hard X-ray component present in FSRQ is usually interpreted as due to
inverse Compton emission, most likely due to a combination of synchrotron
self-Compton emission (with the same population of relativistic electrons
producing the synchrotron radiation and then scattering them to higher
energies) and Comptonization of external radiation (possibly emitted by
material being accreted by the central object; see e.g., Sikora, Begelman \&
Rees 1994).  As the hard X-ray emission in our sources has a similar, flat
slope, it seems natural to attribute it to the same emission mechanism.  The
smaller effect of Doppler boosting for lobe dominated sources would then make
this component to appear and become dominant only at high energies, and this
is exactly what is shown by our data.  This is further confirmed and clearly
shown in Figure 2, where we plot the $2 - 10$ keV spectral index {\it versus}
the core dominance parameter $R$. (The open triangles indicate the SSRQ and
BLRG found in the literature, in some of the papers listed above for FSRQ). It
is evident from the figure that no correlation is present between the two
quantities. This seems in disagreement with the correlation claimed by Lawson
et al. (1992) (based on EXOSAT/ME data) and Lawson \& Turner (1997) (based on
{\it Ginga} data), the only previous hard X-ray studies which included some
steep-spectrum radio quasars. We note that the spectral indices derived from
EXOSAT data had relatively large errors and that our sample of FSRQ, SSRQ and
BLRG is larger than the samples used by Lawson et al. (1992) and Lawson \&
Turner (1997) (26 vs. 18 and 15 objects respectively) especially as far as
lobe-dominated broad-line sources are concerned (where we have basically
doubled the number of available
sources). Moreover, the correlation claimed by Lawson \& Turner (1997)
becomes significant only by excluding the three BLRG in their sample (the
exclusion of the BLRG has no effect on the lack of correlation between
$\alpha_{\rm x}$ and $R$ in our sample). Larger homogeneous samples
(especially of FSRQ) are clearly required in order to investigate this issue
in more details. 

\begin{figure}
\centerline{\psfig{figure=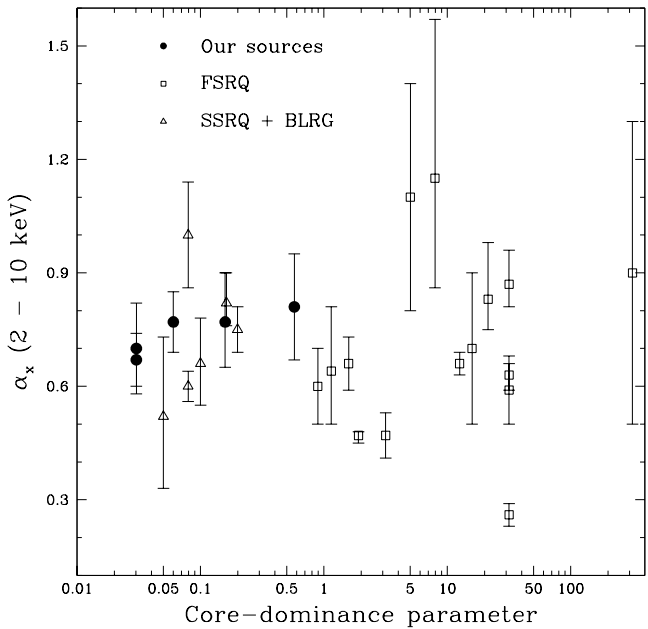,width=9cm,angle=0}}
\caption{The $2 - 10$ keV ($2 - 18$ keV for 10 sources from Lawson \&
Turner 1997) X-ray spectral index versus the core-dominance parameter for our
sources (filled symbols) and core-dominated flat-spectrum radio quasars (FSRQ:
empty squares) and lobe-dominated steep-spectrum radio quasars and broad-line
radio galaxies (SSRQ + BLRG:
triangles) from the literature. Error bars represent 90\% errors for our
objects and 1 $\sigma$ errors for most of the other sources. See text for
details.}
\end{figure}

Our hard X-ray spectra are well fitted by a single power law and we find no
evidence for the hard excess often seen in low-luminosity AGN (e.g., Nandra \&
Pounds 1994) and interpreted as due to Compton reflection of the X-rays off
optically thick material (Guilbert \& Rees 1988; Lightman \& White 1988). In
the case of Pictor A, this is also confirmed by the results of Eracleous \&
Halpern (1998) based on a longer {\it ASCA} observation. We note, 
however, that
this reflection component should normally be apparent above $\sim 10$ keV and
that our data reach these energies only for Pictor A (and even then with
relatively small statistics; see Sect. \ref{sec:pds}).

Wo\'zniak et al. (1998) have recently studied the X-ray (and soft
$\gamma$-ray) spectra of BLRG using {\it Ginga}, {\it ASCA}, OSSE and EXOSAT
data. Their object list includes 4 lobe-dominated BLRG, namely 3C 111, 3C 382,
3C 390.3, and 3C 445. The X-ray spectra have an energy index $\alpha_{\rm x}
\sim 0.7$, with some moderate absorption. Fe K$\alpha$ lines have also been
detected with typical equivalent widths $\sim 100$ eV. Any Compton reflection
component is constrained to be weak and is unambiguously detected only in 3C
390.3. Our results are consistent with their findings. 

Our MECS results for PHL 1657 are in agreement with the {\it Ginga} energy
slope obtained by Williams et al. (1992: see their Table 3), while our $2 -
10$ keV flux appears to be $\sim 35\%$ smaller. Eracleous \& Halpern (1998)
reported on a $\sim 65$ ks {\it ASCA} observations of Pictor A. 
Our LECS and MECS data appear to require a slightly flatter spectral index
than given by these authors ($0.77\pm0.03$, from the SIS and GIS fits) while 
our
$2 - 10$ flux is similar to that derived from the SIS (but $\sim 10\%$ smaller
than the $2 - 10$ keV flux estimated from the GIS). 

As discussed in the Introduction, various previous studies had found that SSRQ
displayed a steep soft X-ray spectrum (see, e.g., Fiore et al. 1998). In fact,
despite the hard component at higher energies, we nevertheless observe a
steeper spectrum at lower energies. In the whole ROSAT band we find
$\alpha_{\rm x} = 1.19\pm0.13$ (excluding OM$-$161, for which the ROSAT
\ax~has large uncertainties), which is intermediate between the values
obtained for SSRQ by Fiore et al. (1998) between $0.4-2.4$ keV ($\alpha_{\rm
x} = 1.14$) and $0.1-0.8$ keV ($\alpha_{\rm x} = 1.37$). Our best fits to the
whole $0.1 - 10$ keV range indeed {\it require} a spectral break 
$\Delta\alpha_{\rm x} \sim 0.5$ between the soft and hard energy slopes at 
about $1 - 2$
keV. The dispersion in the energy indices is larger for the soft component. We
find $\sigma({\alpha_{\rm S}}) = 0.28$ while $\sigma({\alpha_{\rm H}}) =
0.10$, which might suggest a more homogeneous mechanism at higher energies. We
note that Fiore et al. (1998) also found a concave spectrum
($\alpha_{\rm 0.1-0.8~keV} - \alpha_{0.4-2.4~keV} \simeq 0.2$) for radio-loud
AGN (both flat- and steep-spectrum) in the ROSAT band.

There are some concerns (R. Mushotzky, private communication) of
miscalibration between ROSAT, on one side, and {\it BeppoSAX}, {\it ASCA} and 
RXTE on
the other side, which could affect some of our conclusions. Namely, the
inferred ROSAT spectral indices might be steeper than those derived, in the
same band, by other X-ray satellites (a detailed comparison of simultaneous 
{\it ASCA}/RXTE/\sax spectra of 3C 273 is given by Yaqoob et al. in 
preparation). The 
spectral breaks we find in the
spectra of our objects could then be partly due to this effect. This is
clearly an important point, very relevant for X-ray astronomy, but which goes
beyond the scope of this paper. Nevertheless, we can still comment on this as
follows: 1. the ``\sax only'' spectrum of PHL 1657 shows, by itself,
significant evidence of a break, with best fit parameters consistent (within
the rather large errors) with those obtained from the full ROSAT and \sax fit
(see Sect. \ref{sec:lecsmecs}). At least in this source, then, the evidence for
a spectral break is ``ROSAT independent.'' The fact that this is not the case
for the two other objects with LECS data, Pictor A and OM$-$161, can be
explained by the relatively smaller break in the first object and the small
LECS statistics in the latter. In other words, the available evidence is {\it
consistent} with breaks similar to those derived from the combined ROSAT and
\sax fits to be present also in the LECS/MECS data; 2. our main result, that
is, the presence of a hard X-ray component in all our sources at $E \ga 1 - 2$
keV, is based on \sax data and therefore clearly independent of any possible
ROSAT miscalibration.

One could also worry about possible miscalibrations between different X-ray
instruments affecting the (lack of ) correlation in Fig. 2. However, the
results of Wo\'zniak et al. (1998) appear to exclude that possibility. The
{\it ASCA}, {\it Ginga} and EXOSAT X-ray spectra of the sources studied by
these authors, in fact, agree within the errors, particularly in the hard
X-ray band. Given the good cross-calibration between \sax and {\it ASCA},
a large miscalibration between {\it BeppoSAX}, {\it Ginga} and EXOSAT (the 
instruments used to obtain the data used in Fig. 2) seems to be ruled out. 

\section{Conclusions}

The main conclusions of this paper, which presents \sax data for a well
defined sample of 2-Jy steep-spectrum radio quasars and broad-line radio
galaxies can be summarized as below.

All five lobe-dominated, broad-line sources included in this study have been
clearly detected up to 10 keV (50 keV for Pictor A) and display a flat X-ray
spectrum ($\alpha_{\rm x} \sim 0.75$) in the $2 - 10$ keV range. One source
(out of the three with LECS and MECS data, i.e., reaching down to 0.1 keV)
shows significant evidence of a spectral break at $E \sim 1$ keV. When ROSAT
PSPC data, available for all five sources, are included in the fit, the
evidence for concave overall spectra, with $\alpha_{\rm soft} - \alpha_{\rm
hard} \sim 0.5$ and $E_{\rm break} \sim 1.5$ keV, becomes highly significant
for all objects with good enough statistics at low energies (i.e., excluding
OM$-$161). No iron lines are detected in our spectra but the upper limits we
derive are not very stringent (due to the relatively short exposure times).
The flat high-energy slope we find for our lobe-dominated sources is
consistent with the hard X-ray emission present in core-dominated radio
quasars. In fact, by collecting data from the literature on the X-ray spectra
of radio quasars, we show that the available data are consistent with {\it no
dependence} between the $2 - 10$ keV spectral indices and the
core-dominance parameter, somewhat in contrast with the situation at lower
energies. Finally, a thermal emission component is present at low energies in
the spectra of the two broad-line radio galaxies, although only at the
$\sim 10\%$ level. 

Three more targets have been approved as part of this \sax observing program
(one at a lower priority). We will be presenting results on these additional
objects, and a more thorough discussion of the implications of our results in
terms of emission processes, orientation, and more generally unified schemes
in a future paper.

\section*{Acknowledgements}

We acknowledge useful discussions with Alfonso Cavaliere, Andrea Comastri,
Fabrizio Fiore, Paolo Giommi, Paola Grandi, Richard Mushotzky, Tahir Yaqoob.
We thank Paola Grandi also for her help with the analysis of the PDS data of
Pictor A. This research has made use of the NASA/IPAC Extragalactic Database
(NED), which is operated by the Jet Propulsion Laboratory, California
Institute of Technology, under contract with the National Aeronautics and
Space Administration.


\end{document}